\documentclass[12pt]{article}

\usepackage{amstext,amsmath,amssymb,amsfonts,bbm,slashed}
\usepackage[latin1]{inputenc}
\usepackage{epsfig}
\usepackage{hyperref}
\usepackage{amsthm}
\usepackage{subfigure}
\usepackage{color}
\usepackage{multirow}

\usepackage{cancel}

\newcommand{\bea}{\begin{eqnarray}}
\newcommand{\eea}{\end{eqnarray}}
\newcommand{\beq}{\begin{equation}}
\newcommand{\eeq}{\end{equation}}

\begin{document}
\title{Quantum Gravity and Renormalization:\\
The Tensor  Track}
\author{Vincent Rivasseau\\
Laboratoire de Physique Th\'eorique, CNRS UMR 8627,\\
Universit\'e Paris XI,  F-91405 Orsay Cedex, France\\
E-mail:  {\sl rivass@th.u-psud.fr}}

\maketitle

\begin{abstract}
We propose a new program to quantize and renormalize gravity based on recent progress
on the analysis of large random tensors. We compare it briefly with other existing
approaches.
\end{abstract}

\begin{flushright}
LPT-20XX-xx
\end{flushright}
\medskip


\section{Introduction}

Quantization of gravity is still in debate. 
Nature keeps us away from direct experiments. Mathematics gives only a few hints. 
The key problem seems the ultraviolet fluctuations of space-time
near the Planck scale. They cannot be renormalized through 
traditional field theoretic methods. 

Risks are enormous in attacking this problem, among which
mathematical hyper-sophistication is perhaps not the least. Existing approaches 
have been developed for decades, intimidating {\it{de facto}} the newcomer. 
Nevertheless we propose here still another road to quantize gravity. 

It is born of closely related programs namely group field theory (GFT) \cite{boul,laurentgft,Oriti:2011jm}, non commutative quantum field theory (NCQFT) 
\cite{Rivasseau:2007ab}, matrix models \cite{Di Francesco:1993nw} and dynamical triangulations \cite{ambjorn-book,scratch}, and also of many 
discussions with members of the loop quantum gravity (LQG) \cite{Rovelli} community. But it is now sufficiently distinct from all of
these to deserve a name and a presentation of his own. For the moment let us 
call it \emph{tensor field theory}, or simply TFT.

The program is based on a simple physical scenario and on a new mathematical tool.

The physical scenario, called {\emph{geometrogenesis}} or emergence of space-time, is certainly not new: it has been discussed 
extensively in various quantum gravity approaches \cite{ambjorn-book,scratch, laurentgft, Seiberg:2006wf, Konopka:2008hp, Oriti:2006ar,Oriti:2007qd,SindoniGFT,Sindoni:2011ej}. 
At the big bang (which could perhaps more accurately be called the big {\emph {cooling}}), space-time would
condense from a perturbative or dilute pre-geometric phase 
to an effective state which is large and geometric. This state is better and 
better described by general relativity plus matter fields as the 
universe cools further. Our particular trend in this scenario is to emphasize
the role of the renormalization group (RG) as the guiding thread 
throughout the pre-geometric phase as well as afterwards
\footnote{In particular we do not consider, as in 
\cite{Seiberg:2006wf} that the emergence of time 
means the end of determinism; we would prefer to think that 
as long as there is a RG flow, there is an (extended) notion of determinism.}.

The new mathematical tool is the recently discovered universal theory of random tensors and their 
associated 1/N expansions \cite{Gurau:2011kk,Gur3,GurRiv,Gur4,Gurau:2011xp}
\footnote{The theory considers general, ie {\emph{unsymetrized}}, tensors and uses a combinatorial device called color
to  follow their indices, hence its initial name of {\emph{colored}} tensor theory \cite{color}.}. It seems sufficiently simple 
and general to likely describe a generic pre-geometric world.

The program consists in exploring this physical scenario with this new tool in the 
most conservative way, that is using quantum field theory
and the renormalization group.
Only the ordinary axiom of locality of quantum field interactions has to be 
abandoned and replaced by its tensor analog.
 
The first step of this program has been performed. Renormalizable tensor models 
have been defined \cite{BenGeloun:2011rc}. They provide random tensors with natural 
quantum field extensions equipped with their genuine renormalization group. They are higher 
rank analogs of the models of Grosse-Wulkenhaar type \cite{Grosse:2004yu, Gurau:2008vd} 
and of their matrix-like renormalization group.
The next steps in TFT consist in a systematic analysis of the models
in this class, their symmetries, flows, phase transitions and symmetry breaking patterns, as was
done for local quantum field theories in the 60's and 70's and was started for NCQFTs in the last decade. 
We hope to discover theories which lead to a large effective world with 
coordinate invariance\footnote{This invariance 
has been studied in (colored) group field theories in \cite{Baratin:2011tg}. See also the interesting remark about renormalization and diffeomorphism
invariance at the end of section II in \cite{laurentgft}.}, hence likely to bear an Einstein-Hilbert effective action.
Since continuum limits can restore broken symmetries (such as rotation invariance), this
hope does not seem completely unrealistic.

Let us precise the relationship of TFT with all its parents.
TFT can in particular include the study of renormalizable GFT models, which are similar to combinatorial models but with an 
additional gauge invariance. It should be relatively easy
to define such renormalizable GFT models by slight modifications of \cite{BenGeloun:2011rc}, since
combinatorial and GFT models share the same 1/N expansion \cite{Gur3}. However
exclusive attention to GFT models and their richer geometric content could be 
premature until we have better understood the simpler combinatorial models and their phase transitions. 
This is why we consider TFT not just a subprogram of GFT.

Matrix models  provided the first controlled example of the geometrogenesis 
scenario \cite{David:1985nj, Kaz}. Shortly thereafter, tensor models  were advocated as their natural 
higher dimensional generalization \cite{ambj3dqg,mmgravity,sasa1}.
But since the 1/N expansion was missing, the program evolved over the years 
into the mostly numerical study of dynamical triangulations.
TFT uses the same concepts than dynamical triangulations but revives the methods through the
emphasis on renormalization and on the 1/N expansion. We hope and expect in the future rich interactions 
between analytic studies and numerical simulations
in the development of TFT. TFT should also benefit as much as possible from all the expertise gained in the study
of matrix models and of NCQFTs, as TFT is a direct attempt to generalize them to higher rank tensors.

The reader wont find here discussion of asymptotic safeness scenarios \cite{Reuter}; although they share with TFT
the use of RG as a main investigation tool, their approach is quite different. But we do
include in the last section brief comments on the relationship between TFT and two other current approaches
to quantum gravity, namely string theory and LQG. For LQG we have included a contribution to the evolving debate
on whether and how to introduce renormalization in this field\footnote{We thank C. Rovelli for his recent invitation to lecture 
in Marseille on this subject.}.

This paper is neither a review nor meant to be technical. It builds upon a previous similar one \cite{Rivasseau:2011xg},
but with many changes incorporated. Indeed the last year has seen significant mathematical progress 
in the line of thought we advocated.

\section{Why TFT?}

Einstein understood gravity as a \emph{physical theory of geometry}. Wheeler summarized 
his theory in the famous sentence: matter tells space how to curve, space tells matter how to move.
General relativity was founded following the principle of {\emph general invariance 
under change of coordinates}.
% and the condition of asymptotic Newtonian weak field limit?

Feynman in an incredibly bold step reduced quantization to {\emph{randomization}. 
Instead of computing a classical deterministic history, quantization \emph{sums} over all possible such histories
with an action-depending \emph{weight}. In modern simplified notations, it means that 
quantum expectations values are similar to statistical mechanics averages, possibly with an additional factor $i$
\bea   <A>  = \frac{1}{Z}  \int A( \phi)  e^{- iS(\phi ) } d \phi,     \;\;  Z =  \int e^{- iS(\phi ) } d \phi
\eea
where $Z$ is a normalization factor.
In the Euclidean formulation, which corresponds physically to a  finite temperature,
even the additional $i$ factor is absent and functional integral quantization
reduces to \emph{probability theory}. 

If we try to join the two main messages of Einstein and Feynman we get the equation
\bea  Quantum \;\; Gravity \ = \  Random\;\; Geometry. 
\eea
There is quite a wide agreement on this equation among all the main 
schools on quantum gravity, although neither string theory nor LQG take it as their starting point. 
The problem is that geometry is rich. Especially in three or 
four dimensions there does not seem to be a unique way to put a canonical probability theory on it.

We can take inspiration from another outstanding idea of Feynman, namely to represent quantum 
histories by {\emph graphs}. We feel that it is perhaps not sufficiently emphasized in the graph theory community
that quantum field theory and Wick's theorem provide a \emph{canonical measure} for graphs by 
counting graphs through pairings of half-lines into lines\footnote{Counting graphs in this way is 
called zero-dimensional QFT by theoretical physicists.}.
Consider for instance the two connected graphs you can create by joining four vertices of coordination 4.
\begin{figure}[h]
 \includegraphics[width=5cm]{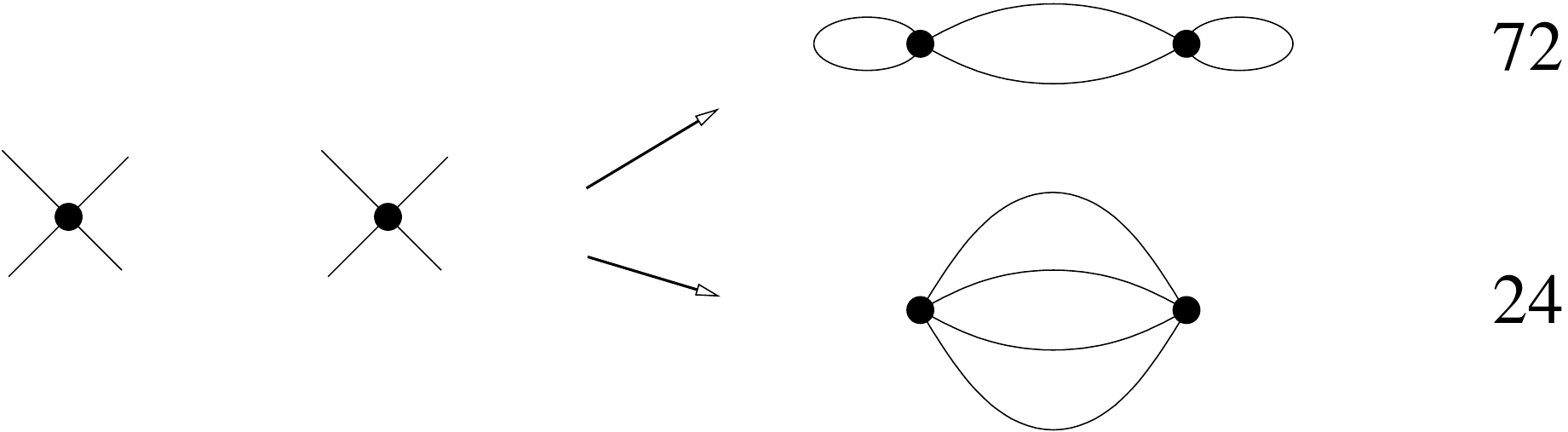}
\end{figure}
The first picture has canonical weight 24. The second picture has canonical weight $72$.
Hence the second picture is three times more probable than the first. 

Graphs are among the simplest pre-geometric objects hence should play a leading role in a theory of
quantum gravity. This was advocated in graphity \cite{Konopka:2008hp}. However there remains a missing link between graphity 
and geometry, which can be borrowed from matrix models and dynamical triangulations. Since
simplicial geometry is dual to stranded graph theory, we can import the canonical notion of probability theory from graphs 
to triangulations and declare it the correct randomization of geometry at the simplicial level.
Combining graphs according to quantum field theory rules we arrive to random tensors. With the
help of the 1/N expansion, the canonical interactions of these random tensors were identified in \cite{Gurau:2011tj}. Finally
adding a natural non trivial propagator to allow for dynamics and renormalization we arrive at TFT  \cite{BenGeloun:2011rc}.

We hope that after having fixed the quantization of this most basic level of geometry, more complicated
geometric structures such as differential and Riemannian aspects will follow
from the renormalization group. Expressed the other way around we could consider
the Planck scale as the limit at which local, continuum physics ceases to apply and 
has to be substituted by tensor physics. In this picture tensor physics is the
more fundamental level, just as QCD is more fundamental than nuclear physics.

%This would certainly not be more surprising than the complexity of chemistry and life
%emerging from bound states built on the simple quantum rules of atomic physics. 

The 1/N expansion provides the tensor program with a new analytic tool.
We could wonder how general is this tool and address now this question.

\section{1/N Expansion and universality}

There is a strong link between the universal character of the central limit theorem in probability theory
and the existence of a 1/N expansion in corresponding field theoretic perturbative expansions.
The initial discovery of the central limit theorem predates by more than a century the invention of the corresponding
vector 1/N expansion in field theory. For random matrices universality of the Wigner-Dyson laws \cite{Mehta} 
was followed rather soon by the 1/N expansion of 't Hooft and followers \cite{Hooft}. But in the case of tensors in fact the 1/N field theoretic 
method slightly predates universality! We have now three versions of the central limit theorem, for independent identically 
distributed (iid) vectors, matrices and tensors. All three are closely related to the corresponding 1/N expansion. This has been 
clarified in \cite{Gurau:2011kk}, to which we refer for details. Here we just provide a sketch of this relationship.

The 1/N expansion is a field theoretic method to find the leading stranded Feynman graphs at fixed perturbative order
which arise in the perturbation expansion of an interacting  quantum field theory. Such a theory
incorporates natural interactions perturbing a Gaussian measure. So it seems at the beginning a very
particular type of non-Gaussian measure.

A theorem of Kolmogorov essentially relates any probability law to its moments, that is to polynomial expectation values.
Hence convergence to Gaussianity in the central limit theorem can be done by studying convergence of 
polynomial moments, or of their cumulants (connected parts). 
Consider large random vectors of size N, matrices of size $N^2$ or tensors of size $N^3$ and so on.
Their coefficients are iid with a fixed \emph{atomic law}. We can restrict easily to an even centered
such atomic law. What distinguishes matrices or tensors
from just longer vector-like lists of variables is that we are not interested in the same \emph{observables}. To take into account 
covariance or contravariance under change of basis,
we are interested only in the statistical behavior  at large N of certain \emph{trace invariants}. 
Universality then reduces to prove that in the expectation value of any fixed 
such cumulant, any atomic moment of order 4 or more is washed out at large N. Hence only the second 
moments of the atomic law survive and influence the value of the large N limit distribution. 

The proof follows from a graphic representation. Any four moment of the atomic distribution (or higher) is an unlikely coincidence
in the cumulant expansion of any moment of the tensor in the large N limit. It turns out to be washed out
for the same reason than the 1/N expansion computes the leading graphs in the corresponding 
field theoretic problem. Hence the 1/N expansion can be considered a grinding machine to erase
information from higher moments in atomic laws.

Because of this universality, we know that we are on solid ground if we base the pre-geometric phase
of quantum gravity on the 1/N tensor-expansion. Even if  at extreme trans-Planckian scales the 
bare atomic laws for the very large tensors that in TFT would describe our
"pre-universe" remain forever unknown, even if they are discrete, such as throwing 
coins or dices, we should end up with the same limiting flow.

But in physics there is another, still richer, notion of universality than the central limit theorem.
It is the RG theory of phase transitions and of their critical indices. 
We know that the RG is another powerful grinding machine which erases information and creates universality.
Only a few marginal or relevant operators can emerge out of it. 
To close the loop we shall now show that RG is also intimately linked 
to 1/N expansions and central limit theorems. In particular it exhibits a parallel hierarchy
of scalar, vector, matrix and tensor types\footnote{Except for the fact that there is no 1/N expansion for scalars.}. 
TFT interprets this hierarchy as that of the increasingly difficult quantum gravity problem in zero, 
one, two and more than two dimensions.

\section{Renormalization}

\subsection{Essential Features}

The essential features needed for renormalization are an action $S$, a scale decomposition,
a notion of locality and a power counting theorem. As we shall see
the notions of scale and locality can be adapted to quite exotic contexts (vectors, matrices, tensors).

The scale decomposition separates the 
ultraviolet ("high fluctuation") scales of the fluctuation fields from the infrared 
scales of the "background field". At fixed attributions for the scales of the lines of a graph, 
some subgraphs play an essential role. They are the \emph{connected subgraphs
whose  internal lines all have higher scale index than all the external lines of the subgraph}. 
Let's call them the "high" subgraphs. 

The renormalization recipe is to check that the divergent high subgraphs
satisfy the locality requirement. If by power counting their local divergent parts are of the
form of the initial action, then the theory is renormalizable.

In ordinary scalar just renormalizable theories such as $\phi^4_4$ the power counting
is summarized in a formula for the divergence degree $\omega$ such as
\bea  \omega = 4 - N  \label{scalar}
\eea
where $N$ is the number of external legs.

Renormalization does not "pull
infinities under the rug" but has a deep physical meaning. 
An external observer (which has only access to low momenta) 
has no choice but to measure effective constants which are the sum of bare constants
plus high momentum radiative corrections.
The parameters of the model change with scale but not the structure of the model itself. 
Renormalization generates a flow between the bare action and renormalized or effective action
which depends on the observation scale.
This flow is computed recursively, step by step, by the renormalization group. 

\subsection{Renormalization Group}

Let us start by a classic citation \cite{Wilson}:

\emph{
The renormalization group approach is a strategy for dealing with problems
involving many length scales. The strategy is to tackle the problem in steps, one
step for each length scale. In the case of critical phenomena, the problem, technically, 
is to carry out statistical averages over thermal fluctuations on all size
scales. The renormalization group approach is to integrate out the fluctuations
in sequence starting with fluctuations on an atomic scale and then moving to
successively larger scales until fluctuations on all scales have been averaged.} 

It is now recognized that RG
governs the standard model, hence all known physics\footnote{ Except perhaps quantum gravity; but quantum gravity
is still unknown physics and TFT precisely postulates that RG also applies to it.}.
RG is the only known way to understand and organize logically 
divergencies in quantum field theory, statistical mechanics and condensed matter.

An elementary step to compute the RG flow is made of two basic sub-steps.
A functional integration over a slice of "fluctuation fields" $\phi_f$ 
is followed by he computation of a logarithm to define the effective action $S' (\phi_b)$
for the background field
\bea I (\phi_b )= \int d\phi_f  e^{- S (\phi_f + \phi_b)} = e^{-S' (\phi_b)}  \to S' (\phi_b) = -\log I (\phi_b) .
\nonumber \eea

The RG  flow is non trivial because these two sub-steps do not commute\footnote{A rescaling step is often added but this is technical rather
than fundamental.}.
An essential feature of this flow is that it is directed. Scales and locality 
match this orientation of the RG arrow, and cannot work the other way around. 
In this way there is a deep analogy between the renormalization 
group and the second law of thermodynamics. The RG flow is directed and irreversible, because 
its coarse graining erases information. Ultimately it depends on an external observer 
who must average with a probability law over the data he cannot access.
The key physical analogies between time, scale and temperature are at the core of 
the paradigm to view the evolution of the universe as a RG cooling trajectory.

\subsection{A hierarchy of Renormalization Group and 1/N Expansions}

We saw already that there is a parallel between the hierarchy of central limit theorems in probability theory 
and the hierarchy of 1/N expansion expansions in quantum field theory. 
There is also an associated hierarchy of renormalization group types: scalar, vector,
matrix, tensors. They can be distinguished by their different notions of locality 
and the different power counting formulas to which they lead to.

We have already discussed scalar-type RG: ordinary quantum field theories such as $\phi^4_4$ or Yang Mills theories are in this category.
In the just renormalizable  case, divergence degrees are simply formulas in the number of external legs such as \eqref{scalar}.

Vector-type renormalization group occurs e.g. in condensed matter when approaching the Fermi surface \cite{FMRT}.
The most divergent graphs are not the 2 and 4 point functions but the 2 and 4-point functions made of
bubble chains at zero external momenta. They clearly form linear chains, hence are associated
to one dimensional spaces, or one-dimensional quantum gravity from the point of view of geometrogenesis.

The epitome for matrix-type renormalization group is the Grosse-Wulkenhaar model.
Here the degree of divergence is given by a formula such as
\bea  \omega = 4- N  - 8 g - 4( B-1)
\eea
where $g$ is the genus and $B$ the number of faces broken by external legs.
Only planar 2 and 4 point graphs whose external legs all arrive on the same external face
do diverge. They satisfy the locality principle (also called Moyality) which simply states
that the matrix equivalent of local operators are traces of powers of the matrix.

In tensor-like renormalization group the formulas are still more complicated and involve sums
over genera of jackets of the graph and of its boundary \cite{BenGeloun:2011rc}. The locality
principle follows again from the identification of the right trace invariants in \cite{Gurau:2011tj}.

We know the renormalization group type can change along a given RG trajectory at a phase transition point.
For instance at the BCS transition in condensed matter, the RG type changes form vector to scalar.
There is therefore no reason the RG cannot change from tensor to lower-rank type at
geometrogenesis.

\subsection{The Case for the Phase Transition}
\label{phasetrans}

Can we escape the geometrogenesis phase transition along the tensor  program? The short answer is no, for
several compelling reasons.

First of all, phase transitions and the formation of bound states is perhaps the most general and generic aspect of all physics.
No interacting physical system when developed over many scales is free of such transitions; for instance in condensed matter
even initially repelling electrons do form bound states. It would be extremely strange, even
almost unbelievable if the theory of quantum gravity was the only physical theory not to exhibit such a phenomenon.

Then two additional arguments apply more specifically to the tensor  program. The first 
has been forcefully argued in eg \cite{Oriti:2007qd}. In any GFT or in any tensor theory that tries to build space-time
out of triangulations, the initial vacuum  is not a particular space but no space at all. 
Hence it cannot be the final product, namely the large universe in which we live. This means there has
to be a phase transition along the way.  

A more mathematical argument relies again on the renormalization group. Only phase transitions
allow a change in the RG power counting and type. But we need such a change
to solve the apparent contradiction between the perfectly renormalizable
pre-geometric theory that we envision as fundamental and the  Einstein-Hilbert action on flat space, well-known to be non-renormalizable
in the ultraviolet regime.
In a phase transition the expansion point in the action changes, hence also the Hessian around this point. It means that the propagators,
the particles and statistics after the transition can be completely different than before. For instance in the BCS phase transition
we go from a Fermionic renormalizable model with a vector-type RG (around the Fermi surface)
to a Bosonic model (the Cooper pair) with a scalar-type RG. We need geometrogenesis
for the initial tensor-type RG to morph into a lower-rank RG types
that will govern the more ordinary flows of matter and gravity (i.e. the local metric field) 
after space-time has developed.

The phase transition scenario is already seen both in matrix and in the simplest tensor models analyzed so far
\cite{David:1985nj,Bonzom:2011zz}.

A last word about trans-Planckian physics. Gravity could have a true and absolute physical cutoff at the Planck scale. 
But most physicists agree that quantum gravity effects start around the Planck scale.
If there is also an absolute UV cutoff at Planck scale there would be no real regime for quantum 
gravity to develop, except in an extremely limited sense of a few scales. If there is a trans-Planckian regime
with many scales we think it should be appropriately called the quantum gravity regime, although
it would be pre-geometric rather than geometric. 

\section{Other Approaches}

\subsection{String Theory}

String theory is a world of its own. 
Lauded as the ultimate theory of everything it has also attracted criticism 
(see \cite{Rovelli:2011mu} for a recent example).  We cannot even try to scratch this subject here.
However since it remains by far the main stream approach to quantum gravity today we 
nevertheless include a few brief
and casual remarks on the most likely possible contact points between string theory 
and the tensor  program. It seems that there are at least two main such contact points, namely dualities
and the AdS-CFT/holography correspondence.

Dualities lie at the heart of string theory but also play an essential role 
in NCQFT. The Grosse-Wulkenhaar model is based on Langmann-Szabo duality,
which seems responsible for its conformal fixed point. We should study which dualities
also occur in  which tensor models.

The AdS-CFT correspondence shows how string theory can reduce to quantum field theory on a boundary,
and the holography principle extends this to any bulk theory of quantum gravity.
The idea remains mostly limited to supersymmetric models. 
TFT should certainly benefit from this beautiful circle of ideas, for instance from the possibility of identifying
the radial direction in AdS-CFT with the RG scale.
There are some preliminary glimpses of a possible holographic nature of the boundary of colored 
tensor graphs.

Another more superficial similarity between tensors and strings is the non local character of their interactions.
Non-locality in string interactions is due to their extended
character, whether the non-locality of tensors looks simpler but more abstract and radical, corresponding
to the gluing rules of simplicial geometry.

String theory now includes branes and just as matrix models quantize the string world sheet, one can hope
that tensor models can provide an appropriate quantization of the branes world volume.

Other superstring features such as  supersymmetry or the Kaluza-Klein/Calabi-Yau ideas
seem quite orthogonal to the tensor  program.
Supersymmetry of course could be included in tensor models but its key property 
to soften or cancel divergencies is not a prerequisite for TFT. 
Calabi-Yau compactification lead to fascinating 
mathematics. But tailoring a particular Calabi-Yau space so friendly to superstrings that moving in it 
they generate the standard model seems truly orthogonal to TFT.

String theory main successes relied on the spectacular development 
in the second half of the XXth century of key mathematical tools to analyze two-dimensional physics: conformal theory, integrability
and random matrices. It is likely that the generalization of these tools to higher dimensions will bring new perspectives. 
Our hope is for the grinding machine of the renormalization group 
to provide a sturdy rather than elegant universe. 

\subsection{Loop Quantum Gravity and Renormalization}

Several reviews discuss the pros and cons
of LQG and spin foams \cite{Rovelli:2011mu,Nicolai, Thiemann, Roche} but they do not focus on the specific question of 
renormalization. The development of GFT and TFT  as more independent programs requires 
to clarify which of their aspects are compatible with LQG (see \cite{Baratin:2011kr} for a related discussion).

LQG, in its covariant version, expresses physical quantities through spin foam amplitudes,
which contain sums over the discrete values $j$ of spin representations of $SU(2)$, or of another related Lie group.
A basic tenet in LQG is that there is no external background geometry, hence the 
Lie group is ``all you have". All known spin foam models, starting with the simplest one
(the Ponzano-Regge spin foams in three dimensions) contain divergencies; the high spin sums
do not always converge. The corresponding power counting is now understood in some detail \cite{FreiGurOriti, sefu3, BS3}.

In LQG high spins are interpreted as large geometries, hence the large spin divergencies are called infrared divergencies.
The discretization at small j's $j=0, 1...$ is interpreted as an  ultraviolet cutoff on the theory at the Planck scale. 
Ultraviolet finiteness  in LQG is hailed, together with discretization of lengths and areas, as a major result of the theory.}
Also geometrogenesis is not part of mainstream LQG; the preferred cosmological scenario is a bounce,
based on the analysis of models with a few degrees of freedom.

We claim that he divergencies of LQG cannot be \emph{renormalized} without major changes to this picture,
for  two main reasons.

First in LQG as it exists today, there is no widely agreed action relating all spin foams.
LQG is therefore only a first quantized theory of gravity. The list of spin foams to sum and their 
exact combinatoric weights for a given computation have to be fixed before any RG treatment. 
Of course individual divergent amplitudes can be regularized 
in many ways to produce finite answers. For instance they can be omitted 
(regularized to 0). But this is definitely not renormalization.

Second, the direction of the RG cannot be changed at will.  
As explained above, \emph{the RG flow is always directed from ultraviolet towards infrared.}
RG averages and erases information on fine details of the theory, not the converse.
In LQG as in GFT there is no background, the Lie group is all one has to sum upon. 
So it is only on that Lie group that one can average in order to renormalize.
RG can then only flow from large (ultraviolet) spins to smaller spins, not the converse.
Locality, even in the extended tensor sense, does not work the other way around \cite{BenGeloun:2011rc}.

On the more positive side, any quantum gravity theory should make contact with physics, and 
produce an effective world which might include aspects of LQG. In particular geometrogenesis
in TFT or GFT should occur at a still giant value of the tensor momenta or spins $j$, 
to account for the enormous number of degrees of freedom that exist in our universe. 
Models of spin foams or GFTs that incorporate the geometric data of general relativity as large $j$ asymptotics
\cite{Muxin,EPR,FK, Baratin:2010wi,Baratin:2011hp}
might therefore be very important to understand how 
geometrogenesis leads to our world. 

In short the issue of ultraviolet versus infrared in LQG is not just about names.
To include renormalization, LQG must first adopt a second-quantized formalism such as GFT\footnote{We are still far from this situation.
Although spin foams were rather early identified as Feynman graphs \cite{Freidel:1999jf,Reisenberger:2000fy} and GFT was revived
by the LQG community \cite{laurentgft}, surprisingly LQG has still not adopted GFT as its proper foundation, hence remains today incomplete.}. 
Even then, divergencies in LQG can be renormalized
only as ultraviolet divergencies. This has far-reaching consequences 
such as geometrogenesis\footnote{Geometrogenesis
suggests that bounce is unlikely but cannot rule it out completely for the current big-bang. Something could have heated 
the universe close to the boiling point and it could have re-cooled.}. 

%In short the alternatives for LQG are either to change or to renounce renormalization. 

\section{Conclusion}

The directed renormalization group flow is the modern avatar of
the second law of thermodynamics. Therefore I would like to conclude this paper 
by paraphrasing the famous words of Sir Arthur Eddington in 
"The Nature of the Physical World" (1927):

\vskip.2cm
\emph{
The flow of the renormalization group holds, I think, the supreme position among the laws of Nature. If someone points out to you 
that your pet theory of the universe is in disagreement with Maxwell's equations - then so much the worse for Maxwell's equations. 
If it is found to be contradicted by observation - well, these experimentalists do bungle things sometimes. But if your theory is found to be against the
flow of the renormalization group I can give you no hope; there is nothing for it but to collapse in deepest humiliation.}

\medskip\noindent
{\bf Acknowledgments}

We thank A. Baratin, J. Ben Geloun, R. Gurau,  D. Oriti and A. Tanasa for fruitful discussions and for reading the draft of this paper.
We also thank the organizers of the VIIIth International Conference on Progress in Theoretical Physics in Constantine
that led to this publication.


\begin{thebibliography}{99}


  %\cite{Oriti:2011jm}
  
 
\bibitem{boul}
D. V. Boulatov, ``A Model of three-dimensional lattice gravity,"
Mod. Phys. Lett. {\bf  A7} (1992) 1629;
{\tt hep-th/9202074};

\bibitem{laurentgft}
  L.~Freidel,
  ``Group field theory: An overview,''
  Int.\ J.\ Theor.\ Phys.\  {\bf 44}, 1769 (2005)
  [arXiv:hep-th/0505016].
  %%CITATION = IJTPB,44,1769;%%
  
\bibitem{Oriti:2011jm}
D.~Oriti,
``The microscopic dynamics of quantum space as a group field theory,''
  arXiv:1110.5606 [hep-th].
  %%CITATION = ARXIV:1110.5606;%%


%\cite{Rivasseau:2007ab}
\bibitem{Rivasseau:2007ab}
  V.~Rivasseau,
 ``Non-commutative renormalization,'' in "Quantum Spaces",
Progress in Mathematical Physics {\bf 53}, Birkh\"auser, 2007
  arXiv:0705.0705 [hep-th].
  %%CITATION = ARXIV:0705.0705;%%
  
  \bibitem{Di Francesco:1993nw}
  P.~Di Francesco, P.~H.~Ginsparg and J.~Zinn-Justin,
  ``2-D Gravity and random matrices,''
  Phys.\ Rept.\  {\bf 254}, 1 (1995)
  [arXiv:hep-th/9306153].
  %%CITATION = PRPLC,254,1;%%
  
\bibitem{ambjorn-book}
  J.~Ambjorn, B.~Durhuus and T.~Jonsson,
  ``Quantum geometry. A statistical field theory approach,''
%\href{http://www.slac.stanford.edu/spires/find/hep/www?irn=4205618}{SPIRES entry}
{\it  Cambridge, UK: Univ. Pr., 1997. (Cambridge Monographs in Mathematical Physics). 363 p}


\bibitem{scratch}
R. Loll, J. Ambjorn, J. Jurkiewicz,
The Universe from Scratch,
hep-th/0509010,
Contemp.Phys. 47 (2006) 103-117

  
\bibitem{Rovelli}
Carlo Rovelli, Quantum Gravity, Cambridge Univ. Press 
(Novembre 2004), ISBN 0-521-83733-2.

%\cite{Seiberg:2006wf}
\bibitem{Seiberg:2006wf}
N.~Seiberg,
``Emergent space-time,''
arXiv:hep-th/0601234.
  %%CITATION = HEP-TH/0601234;%%
 
\bibitem{Konopka:2008hp}
  T.~Konopka, F.~Markopoulou and S.~Severini,
 ``Quantum Graphity: a model of emergent locality,''
  Phys.\ Rev.\  D {\bf 77}, 104029 (2008)
  [arXiv:0801.0861 [hep-th]].

%\cite{Oriti:2006ar}
\bibitem{Oriti:2006ar}
D.~Oriti,
``A quantum field theory of simplicial geometry and the emergence of space-time,''
J.\ Phys.\ Conf.\ Ser.\  {\bf 67}, 012052 (2007)
[arXiv:hep-th/0612301].
%%CITATION = 00462,67,012052;%%

%\cite{Oriti:2007qd}
\bibitem{Oriti:2007qd}
  D.~Oriti,
``Group field theory as the microscopic description of the quantum space-time
  fluid: a new perspective on the continuum in quantum gravity,''
  arXiv:0710.3276 [gr-qc].
  %%CITATION = ARXIV:0710.3276;%%

\bibitem{SindoniGFT}
L.~Sindoni,
 ``Gravity as an emergent phenomenon: a GFT perspective,''
arXiv:1105.5687

%\cite{Sindoni:2011ej}
\bibitem{Sindoni:2011ej}
L.~Sindoni, ``Emergent models for gravity: an overview,''
  arXiv:1110.0686 [gr-qc].
  %%CITATION = ARXIV:1110.0686;%%

%\cite{Gurau:2011kk}
\bibitem{Gurau:2011kk}
  R.~Gurau,
  ``Universality for Random Tensors,''
  arXiv:1111.0519 [math.PR].
  %%CITATION = ARXIV:1111.0519;%%

\bibitem{Gur3}
  R.~Gurau,
``The 1/N expansion of colored tensor models,''
  Annales Henri Poincar\'e {\bf 12}, 829 (2011)
  [arXiv:1011.2726 [gr-qc]].
  %%CITATION = AHPJF,12,829;%%

\bibitem{GurRiv}
R.~Gurau and V.~Rivasseau,
``The 1/N expansion of colored tensor models in arbitrary dimension,''
  Europhys.\ Lett.\  {\bf 95}, 50004 (2011)
  [arXiv:1101.4182 [gr-qc]].
  %%CITATION = EULEE,95,50004;%%

\bibitem{Gur4}
  R.~Gurau,
  ``The complete 1/N expansion of colored tensor models in arbitrary
  dimension,''
  arXiv:1102.5759 [gr-qc].
  %%CITATION = ARXIV:1102.5759;%%
  
%\cite{Gurau:2011xp}
\bibitem{Gurau:2011xp}
  R.~Gurau and J.~P.~Ryan,
  ``Colored Tensor Models - a review,''
  arXiv:1109.4812 [hep-th].
  %%CITATION = ARXIV:1109.4812;%%



%\cite{BenGeloun:2011rc}
\bibitem{BenGeloun:2011rc}
  J.~Ben Geloun and V.~Rivasseau,
``A Renormalizable 4-Dimensional Tensor Field Theory,''
  arXiv:1111.4997 [hep-th].
  %%CITATION = ARXIV:1111.4997;%%



  %\cite{Grosse:2004yu}
\bibitem{Grosse:2004yu}
  H.~Grosse and R.~Wulkenhaar,
``Renormalisation of phi**4 theory on noncommutative R**4 in the matrix base,''
  Commun.\ Math.\ Phys.\  {\bf 256}, 305 (2005)
  [arXiv:hep-th/0401128].
  %%CITATION = CMPHA,256,305;%%
  
%\cite{Gurau:2008vd}
\bibitem{Gurau:2008vd}
  R.~Gurau, J.~Magnen, V.~Rivasseau and A.~Tanasa,
  ``A translation-invariant renormalizable non-commutative scalar model,''
  Commun.\ Math.\ Phys.\  {\bf 287}, 275 (2009)
  [arXiv:0802.0791 [math-ph]].
  %%CITATION = CMPHA,287,275;%%
  
  %\cite{Baratin:2011tg}
\bibitem{Baratin:2011tg}
  A.~Baratin, F.~Girelli and D.~Oriti,
``Diffeomorphisms in group field theories,''
  Phys.\ Rev.\  D {\bf 83}, 104051 (2011)
  [arXiv:1101.0590 [hep-th]].
  %%CITATION = PHRVA,D83,104051;%%


\bibitem{color}
 R.~Gurau,
  ``Colored Group Field Theory,''
  Commun.\ Math.\ Phys.\  {\bf 304}, 69 (2011)
  [arXiv:0907.2582 [hep-th]].
  %%CITATION = CMPHA,304,69;%%


%\cite{David:1985nj}
\bibitem{David:1985nj}
  F.~David,
  ``A Model Of Random Surfaces With Nontrivial Critical Behavior,''
  Nucl.\ Phys.\  B {\bf 257}, 543 (1985).
  %%CITATION = NUPHA,B257,543;%%

\bibitem{Kaz}
V.~A.~Kazakov, 
``Bilocal regularization of models of random surfaces,'' 
Phys.\ Lett.\  B {\bf 150}, 282 (1985). 


  \bibitem{ambj3dqg}
  J.~Ambjorn, B.~Durhuus and T.~Jonsson,
  ``Three-Dimensional Simplicial Quantum Gravity And Generalized Matrix Models,''
  Mod.\ Phys.\ Lett.\  A {\bf 6}, 1133 (1991).
  %%CITATION = MPLAE,A6,1133;%%

\bibitem{mmgravity}
 M.~Gross,
``Tensor models and simplicial quantum gravity in $>$ 2-D,''
Nucl.\ Phys.\ Proc.\ Suppl.\  {\bf 25A}, 144 (1992).
  %%CITATION = NUPHZ,25A,144;%%

\bibitem{sasa1}
N.~Sasakura,
 ``Tensor model for gravity and orientability of manifold,''
Mod.\ Phys.\ Lett.\  A {\bf 6}, 2613 (1991).
%%CITATION = MPLAE,A6,2613;%%

\bibitem{Reuter} M. Reuter, F. Saueressig,
 ``Renormalization Group Flow of Quantum Gravity in the Einstein-Hilbert Truncation,''
Phys.Rev. {\bf D65}: 065016, 2002.

%\cite{Rivasseau:2011xg}
\bibitem{Rivasseau:2011xg}
V.~Rivasseau,
``Towards Renormalizing Group Field Theory,''
PoS C {\bf NCFG 2010}, 004 (2010)
[arXiv:1103.1900 [gr-qc]].
%%CITATION = POSCI,CNCFG2010,004;%%

\bibitem{Gurau:2011tj}
R.~Gurau,
``A generalization of the Virasoro algebra to arbitrary dimensions,''
Nucl.\ Phys.\  B {\bf 852}, 592 (2011)
[arXiv:1105.6072 [hep-th]].
%%CITATION = NUPHA,B852,592;%%

\bibitem{Mehta}
M. L. Mehta,  ``Random Matrices,'' Elsevier, 2004 Pure and Applied Mathematics (Amsterdam), 142.

\bibitem{Hooft}
 G. 't Hooft, ``A PLANAR DIAGRAM THEORY FOR STRONG INTERACTIONS,'' Nucl. Phys. B  {\bf 72},
461 (1974).

\bibitem{Wilson}
K. Wilson, ``The renormalization group and critical phenomena,''
Rev. Mod. Phys. 55, 583-600 (1983), Nobel Lecture 1982.

\bibitem{FMRT}
 J. Feldman, J. Magnen, V. Rivasseau and E. Trubowitz, ``An Intrinsic 1/N Expansion for Many Fermion Systems," Europhys. Letters  {\bf 24}, 437 (1993).

%\cite{Bonzom:2011zz}
\bibitem{Bonzom:2011zz}
  V.~Bonzom, R.~Gurau, A.~Riello and V.~Rivasseau,
 ``Critical behavior of colored tensor models in the large N limit,''
  Nucl.\ Phys.\  B {\bf 853}, 174 (2011)
  [arXiv:1105.3122 [hep-th]].
  %%CITATION = NUPHA,B853,174;%%

 %\cite{Rovelli:2011mu}
\bibitem{Rovelli:2011mu}
  C.~Rovelli,
 ``A critical look at strings,''
arXiv:1108.0868 [hep-th].
%%CITATION = ARXIV:1108.0868;%%

\bibitem{Nicolai}
H. Nicolai, K. Peeters, M. Zamaklar,
 ``Loop quantum gravity: an outside view,
arXiv:hep-th/0501114; 
Class.Quant.Grav.  {\bf 22}: R193, 2005.
  
  \bibitem{Thiemann}
 T. Thiemann,  ``Loop Quantum Gravity: An Inside View,''
  arXiv:hep-th/0608210, Lect. Notes Phys.  {\bf 721}: 185-263, 2007.
  
\bibitem{Roche}
S. Alexandrov, P. Roche,
 ``Critical Overview of Loops and Foams,''
arXiv:1009.4475,
Phys.Rept. {\bf 506}: 41-86, 2011.
 
%\cite{Baratin:2011kr}
\bibitem{Baratin:2011kr}
  A.~Baratin and D.~Oriti,
  ``Ten questions on Group Field Theory (and their tentative answers),''
  arXiv:1112.3270 [gr-qc].
  %%CITATION = ARXIV:1112.3270;%%

\bibitem{FreiGurOriti}
  L.~Freidel, R.~Gurau and D.~Oriti,
  ``Group field theory renormalization - the 3d case: power counting of
  divergences,''
  Phys.\ Rev.\  D {\bf 80}, 044007 (2009)
  [arXiv:0905.3772 [hep-th]].
  %%CITATION = PHRVA,D80,044007;%%

%\bibitem{sefu1}
%  J.~Magnen, K.~Noui, V.~Rivasseau and M.~Smerlak,
 % ``Scaling behavior of three-dimensional group field theory,''
 % Class.\ Quant.\ Grav.\  {\bf 26}, 185012 (2009)
 % [arXiv:0906.5477 [hep-th]].
  %%CITATION = CQGRD,26,185012;%%

%\bibitem{sefu2}
%  J.~Ben~Geloun, J.~Magnen and V.~Rivasseau,
%  ``Bosonic Colored Group Field Theory,''
%  Eur.\ Phys.\ J.\  C {\bf 70}, 1119 (2010)
 % [arXiv:0911.1719 [hep-th]].
  %%CITATION = EPHJA,C70,1119;%%

\bibitem{sefu3}
   J.~Ben~Geloun, T.~Krajewski, J.~Magnen and V.~Rivasseau,
  ``Linearized Group Field Theory and Power Counting Theorems,''
  Class.\ Quant.\ Grav.\  {\bf 27}, 155012 (2010)
  [arXiv:1002.3592 [hep-th]].
  %%CITATION = CQGRD,27,155012;%%

 %\bibitem{BS1}
 % V.~Bonzom and M.~Smerlak,
 %``Bubble divergences from cellular cohomology,''
 % Lett.\ Math.\ Phys.\  {\bf 93}, 295 (2010)
 % [arXiv:1004.5196 [gr-qc]].
  %%CITATION = LMPHD,93,295;%%

%\bibitem{BS2}
 % V.~Bonzom and M.~Smerlak,
%``Bubble divergences from twisted cohomology,''
 % arXiv:1008.1476 [math-ph].
 %%CITATION = ARXIV:1008.1476;%%

\bibitem{BS3}
  V.~Bonzom and M.~Smerlak,
 ``Bubble divergences: sorting out topology from cell structure,''
  arXiv:1103.3961 [gr-qc].
  %%CITATION = ARXIV:1103.3961;%%


\bibitem{Muxin}
Muxin Han, Mingyi Zhang,
Asymptotics of Spinfoam Amplitude on Simplicial Manifold: Euclidean Theory,
arXiv:1109.0500.




\bibitem{EPR}
 J. Engle, R. Pereira, and C. Rovelli, ``Loop-quantum-gravity vertex amplitude,''  Phys. Rev. Lett.  {\bf 99},
161301 (2007), arXiv:0705.2388.

\bibitem{FK}
 L. Freidel and K. Krasnov,  ``A new spin foam model for 4d gravity,''  Class. Quant. Grav. {\bf 25},
125018 (2008), arXiv:0708.159.



%\cite{Baratin:2010wi}
\bibitem{Baratin:2010wi}
  A.~Baratin and D.~Oriti,
  ``Group field theory with non-commutative metric variables,''
  Phys.\ Rev.\ Lett.\  {\bf 105}, 221302 (2010)
  [arXiv:1002.4723 [hep-th]].
  %%CITATION = PRLTA,105,221302;%%

%\cite{Baratin:2011hp}
\bibitem{Baratin:2011hp}
  A.~Baratin and D.~Oriti,
  ``Group field theory and simplicial gravity path integrals: A model for
  Holst-Plebanski gravity,''
  arXiv:1111.5842 [hep-th].
  %%CITATION = ARXIV:1111.5842;%%

%\cite{Freidel:1999jf}
\bibitem{Freidel:1999jf}
  L.~Freidel and K.~Krasnov,
  ``Simple spin networks as Feynman graphs,''
  J.\ Math.\ Phys.\  {\bf 41}, 1681 (2000)
  [arXiv:hep-th/9903192].
  %%CITATION = JMAPA,41,1681;%%
  
%\cite{Reisenberger:2000fy}
\bibitem{Reisenberger:2000fy}
  M.~Reisenberger and C.~Rovelli,
  ``Spin foams as Feynman diagrams,''
  arXiv:gr-qc/0002083.
  %%CITATION = GR-QC/0002083;%%


\end{thebibliography}
\end{document}